\documentclass[preprint,preprintnumbers,amsmath,amssymb]{revtex4}
\usepackage{graphicx}

\begin{document}
\draft
Published as: R. Ehrlich, Astropart. Phys.,{\bf 66}, 11-17, (2015).
\\
\title{Six observations consistent with the electron neutrino being a tachyon with mass: $m_{\nu_e}^2=-0.11 \pm 0.016 eV^2$}

\author{ Robert Ehrlich}
\address{George Mason University, Fairfax, VA 22030, USA}
\email{rehrlich@gmu.edu}
\date{\today}

\begin{abstract}
Six observations based on data and fits to data from a variety of areas are consistent with the hypothesis that the electron neutrino is a $m_{\nu_e}^2 = -0.11 \pm 0.016 eV^2$ tachyon.  The data are from areas including CMB fluctuations, gravitational lensing, cosmic ray spectra, neutrino oscillations, and $0\nu$ double beta decay.  For each of the six observations it is possible under explicitly stated assumptions to compute a value for $m_{\nu_e}^2,$  and it is found that the six values are remarkably consistent with the above cited $\nu_e$ mass $(\chi^2=2.73).$  There are no known observations in clear conflict with the claimed result, nor are there predicted phenomena that should occur which are not seen.  Three checks are proposed to test the validity of the claim, one of which could be performed using existing data.\\

\end{abstract}
\maketitle
\section{Introduction}
In two 1999 papers this author suggested the hypothesis that $\nu_e$ was a $m_{\nu_e}^2 \approx -0.25 \pm 0.13 eV^2$ tachyon, based on a pair of  cosmic ray analyses.\citep{Ehrlich1,Ehrlich2}   Here we discuss six observations based on data from CMB fluctuations, gravitational lensing, neutrino mixing, $0\nu$ double beta decay, and cosmic rays that are consistent with that hypothesis, and which yield the revised estimate for the $\nu_e$ mass of $m_{\nu_e}^2= -0.11 \pm 0.016 eV^2$ or equivalently $\mu_{\nu_e} = 0.33 \pm 0.024 eV$ where $\mu_{\nu_e}=\sqrt{-m_{\nu_e}^2}$ -- see a summary in Table I and Fig. 1. The first 1999 claim was based on a model\citep{Ehrlich1} that fit the cosmic ray spectrum, assuming the knee is the threshold for proton beta decay.  As shown by Chodos et al.\citep{Chodos} this process becomes energetically allowed if the electron neutrino is a tachyon and 
\begin{equation}
E_{knee} = \frac{m_p|Q|}{\mu_{\nu_e}}=\frac{1.695 PeV}{\mu_{\nu_e}(eV)}.\end{equation}  
where $m_p$ is the proton mass and Q is the negative Q-value for the decay -- see supplementary animation file.\citep{footnote} 
The model's essential feature was that the decay of cosmic ray protons when $E>E_{knee},$ results in a decay chain: $p\rightarrow n \rightarrow p\rightarrow n \rightarrow \cdots$ that continues until the baryon's energy drops below $E_{knee},$ shifting them to lower energies, and thereby (a) giving rise to the knee, and (b) a pile-up of neutrons just above it, i.e., a small peak at  $E = 4.5 \pm 2.2 PeV.$\citep{Ehrlich1} Neutrons, mostly point back to their sources, unlike protons whose directions are affected by the galactic magnetic field. Thus, if the baryon in the decay chain spends a significant proportion of its time as a neutron, much of its directional information should be preserved.  Moreover, the hypothesized decay chain could allow PeV neutrons to reach us from sources normally considered too distant, given the neutron lifetime.  The second 1999 paper\citep{Ehrlich2} claimed evidence for $6\sigma$ peak centered on $5 PeV$ based on Lloyd-Evans data for Cygnus X-3.\citep{Lloyd-Evans}   Apart from skepticism of this claim, there is also much skepticism about Cygnus X-3 ever being a source of PeV cosmic rays. However, the basis of that skepticism may be poorly justified, especially if Cygnus X-3 is an episodic source, and if a weak $E \approx 4.5 PeV$ signal needs cuts to suppress background, as discussed in detail in Appendix I of ref.~\citep{Ehrlich4}.  Given the nature of the tachyon neutrino hypothesis, however, and the alternative explanations that existed for those earlier cosmic ray analyses the hypothesis has not been taken seriously in the cosmic ray community.  It is therefore important that among the six observations only two involves cosmic ray physics.

\begin{figure}[h]
\includegraphics[width=80mm]{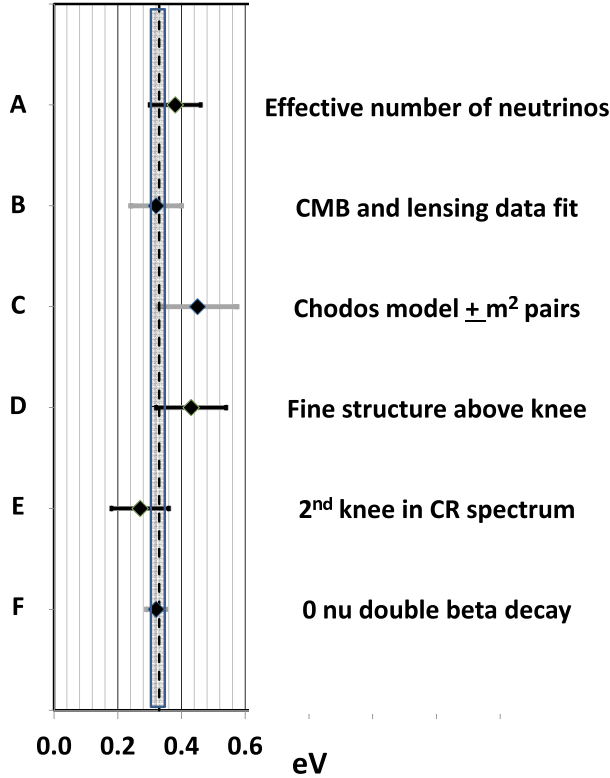}
\caption{A graphical representation of the six tachyonic mass values for $\mu_{\nu_e}$ listed in Table 1.  The error bars, except for E, are based on values from fits to data done by other researchers.  The dashed vertical line and the width of the shaded rectangle show the weighted average mass and its uncertainty.  When the weights are chosen as $w_j = \Delta \mu_j^{-2}/\Sigma \Delta \mu_j^{-2},$ this weighted average corresponds to doing a least squares fit to the six observations to a vertical straight line. Three of the six error bars are shown light grey, which denotes that the inclusion of these items is more conditional than the others: F denotes a disputed experimental result for which the sign of $m_{\nu_e}^2$ is indeterminate, and B and C denote two alternative interpretations of the same cosmological data, only one of which might be true.} 
\end{figure}

\begin{table}
\begin{ruledtabular}
\begin{tabular}{ll}
Published source of data or fits from which value of $\mu_{\nu_e}=\sqrt{-m_{\nu_e}^2}$ is inferred$^*$ & $\mu_{\nu_e}$ in eV\\
\hline
Calculation based on $N_\nu$ by Davies and Moss\citep{Davies} & $0.38 \pm 0.08$\\
Fits to CMB and gravitational lensing data\citep{Battye, Hamann} & $0.320 \pm 0.081$\\
Requirement of Chodos model that neutrinos constitute $\pm m^2$ pairs\citep{Chodos2} & $0.450\pm 0.124$\\
Fine structure seen in cosmic ray spectrum just above knee\citep{rio_conf, Apel, IceTop} & $0.43\pm 0.11$\\
The second knee in the cosmic ray spectrum\citep{Tunka1,Bergman} & $0.27\pm 0.09$\\
Disputed H-M neutrino-less double beta decay experiment\citep{Heidelburg1} & $0.32 \pm 0.03^{**}$\\
\hline
Average of 6 observations and their consistency: $\chi^2=2.73 (p = 74\%)$ & $0.33 \pm 0.024$\\
Average of first 5 observations & $0.36 \pm 0.041$\\
\end{tabular}
\caption{\label{tab:table1}\footnotesize{Tachyonic neutrino mass inferred from six observations.   Although the best fit tritium beta decay results have $m_{\nu_e}^2<0$ they have not been listed here owing to their large uncertainties. The second and third above observations represent two different interpretations of the same CMB and lensing data -- the first assuming no sterile neutrinos, and the second assuming one or more. $^*$The inference of the values listed is that of this author, and not that of the cited works. $^{**}$As discussed in the text, the sign of $m_{\nu_e}^2$ is ambiguous in a double beta decay experiment.}}

\end{ruledtabular}
\end{table}

\section{Observations consistent with $\mu_{\nu_e} = 0.33 \pm 0.024 eV$}

The observations consist of interpretations of published data and fits to those data, and in each case it is possible to compute a value for $\mu_{\nu_e} \pm \Delta \mu_{\nu_e},$ where the $\Delta \mu_{\nu_e}$ are found by error propogation, given the uncertainty in the least well-known quantity on which $\mu_{\nu_e}$ depends.  The consistency of the six observations with the stated mass requires making a variety of explicitly stated assumptions, as discussed in what follows.

\subsection{The effective number of neutrinos and the neutrino masses}
Direct tests of the masses of the neutrinos from particle physics and cosmology usually are only able to set upper limits.  For example, the two most precise experiments measuring the spectrum of tritium beta decay yield $m_{\nu_e}<2 eV,$ \citep{PDG}, while in one recent investigation Davies and Moss (DM) have set an upper limit on the magnitude of the mass of any tachyonic neutrino based on cosmology as $\mu<0.33 eV.$\citep{Davies} 

DM place this upper bound on the neutrino mass using a relation they derive as:

\begin{equation}
\mu  = \sqrt{\frac{2(T^2_{nuc}T^2_{weak})|\Delta N_{\nu}|}{3(T^2_{weak}-T^2_{nuc})}}
\end{equation}

where $\Delta N_\nu = N_\nu-3.05$, $N_\nu$ being the effective number of neutrino species defined in terms of the energy densities of neutrinos and photons at the time of nucleosynthesis, $T_{weak}$ is the temperature of neutrino decoupling, and $T_{nuc}$ is the temperature at the time of nucleosynthesis.  DM actually write Eq. 2 as an upper bound on $\mu$ rather than an actual value, since they use an upper bound on $|\Delta N_{\nu}|$ and not an actual value.  They then use Eq. 2 to cite several values of that upper bound which depend on their assumptions for the parameters.  The more conservative of their two upper bounds is found using data on CMB fluctuations, for which $T_{nuc}$ is replaced by $T_{eq},$ the temperature at matter-radiation equality for which density fluctuations start to grow.  DM use values for $\Delta N_\nu>-0.3,$ $T_{eq}=0.74 eV,$ and $T_{weak}= 0.8 MeV$ from which they obtain $\mu\le 0.33 eV.$  However, if we substitute in Eq. 2 a more up to date value for $\Delta N_\nu = 0.40\pm 0.17,$ based on CMB fluctuations and big bang nucleosynthesis (BBN),\citep{Steigman} we obtain an actual value for $\mu$ rather than an upper bound:  $\mu = 0.38 \pm .08 eV,$ where the uncertainty $\Delta \mu$ is here taken to be $\pm 21\%$ based on the propagation of errors for the $\pm 42\%$ uncertainty in $\Delta N_\nu$ -- the chief source of uncertainty in the DM calculation.  It is also assumed that the neutrino that is the source of dark energy is $\nu_e.$

\subsection{Fits to CMB and gravitational lensing}
A  second observation is based on results from a 2014 article by Battye and  Moss (BM).\citep{Battye}    BM perform fits to five data sets involving the CMB and lensing measurements in order to determine the sum of the 3 active neutrino masses, under two scenarios: 3 + 0, i.e., only three active neutrinos, and 3 + 1, three active and one sterile neutrino.  Very similar fits were reported about the same time by Hamann and Hasenkamp.\citep{Hamann}  Both pairs of authors note that their fits are able to resolve a pair of discrepancies that exists between CMB and lensing data, and they obtain values that are significantly different from zero rather than merely upper limits for the sum of the neutrino masses.  The BM best fit for the case of three active neutrinos only is $\Sigma m_{\nu} = 0.320 \pm 0.081 eV,$ which is about $4\sigma$ from zero.  BM note that given the large value found for $\Sigma m_{\nu}$ compared to much smaller values of $\Delta m^2$ from neutrino oscillation experiments the neutrino masses would need to be nearly degenerate, which would apparently yield $m_{\nu_e} \approx m_{\nu_\mu} \approx m_{\nu_\tau}=\frac{1}{3}(0.320)=0.11 eV.$  Below we discuss an alternative interpretation of their result, which allows for the possibility of some of the neutrino flavors being tachyons.   

The basis of using fits to CMB and lensing data to deduce a value for $\Sigma m_\nu$ starts with the dependence of those data on the spatial energy density of neutrinos $\rho_\nu$ at the time when the CMB fluctuations started to grow.  The overall $\rho_\nu$ can be expressed in terms of the three neutrino flavor masses and their associated number densities: $\rho_\nu=m_{\nu_e}n_{\nu_e}+m_{\nu_\mu}n_{\nu_\mu}+m_{\nu_\tau}n_{\nu_\tau}.$  However, since the number densities should be all equal, given that  the flavors were produced in equal abundance in earlier very high energy interactions, we have $\rho_\nu=n_\nu \Sigma m_\nu.$  It has long been known that tachyons can have a negative energy,\citep{Dhar}, and that negative energy density offers a simple way to explain dark energy,\citep{Nojiri}, one form of which might involve a sea of tachyonic neutrinos.\citep{Davies, Jentschura2}  A negative energy density for tachyonic neutrinos $\rho_\nu=n_\nu m_\nu,$ requires that their mass $m_\nu$ be considered to be negative since their spatial number density, $n_\nu$ cannot be, but note that we are referring here to their gravitational mass, not their kinematic mass, which is of course imaginary.
Given the foregoing, if only the electron neutrino were a tachyon the BM result would need to be written as:

\begin{equation}
\Sigma m_\nu = m_{\nu_\mu} + m_{\nu_\tau} - \mu_{\nu_e} = 0.320 \pm 0.081 eV
\end{equation}

Now, the measured $\Delta m^2$ values from oscillation experiments are between neutrino mass not flavor states, where the relationship between the masses of the two types of states assuming CP conservation can be expressed as: 

\begin{equation}
m^2_{F,i}=\Sigma|U_{i,j}|^2 m^2_j
\end{equation}

and conversely:

\begin{equation}
m^2_{i}=\Sigma|U_{j,i}|^2 m^2_{F,j}
\end{equation}

It was noted earlier that BM used the near-degeneracy of the three mass states as required by the smallness of $\Delta m^2$ relative to $\Sigma m$, to argue that the three flavor states are also nearly degenerate, which is clear from Eq. 4.  However, if one or more of the three neutrino flavors is a tachyon the preceding no longer logically follows.  If we were, however, to assume the ${\emph magnitudes}$ of the three flavor masses are nearly equal, i.e.,  $m_{\nu_\mu} \approx m_{\nu_\tau} \approx \mu_{\nu_e},$ then Eq. 3 would imply: $\mu_{\nu_e} \approx 0.320 \pm 0.08 eV,$ or $m^2_{\nu_e}  = -0.11 \pm 0.05 eV^2.$  Thus, we interpret the BM 3 + 0 fit result as being consistent with there being three active flavor neutrinos of nearly equal magnitude masses, and no sterile neutrinos.  Two of the flavors are assumed to comprise a tachyon-tardyon pair, which is very similar to a recent model by Chodos discussed below.\citep{Chodos2}  As we shall see, there is a simple way for such a large magnitude $m^2_{\nu_e}$ to satisfy the constraints imposed by the small values of $\Delta m^2_{i,j}$ from oscillation experiments. 

\subsection{Chodos model requirement of $\pm m^2$ tachyon-tardyon pairs}
Theories of tachyonic neutrinos have encountered difficulties in terms of satisfying Lorentz Invariance (LI), a matter that some theorists have tried to deal with through possible alternatives to the Lorentz group.\citep{Kostelecky} Of possible relevance is Very Special Relativity (VSR), introduced by Cohen and Glashow, which involves reducing the symmetry to a subgroup of the Lorentz Group that produces nearly all the consequences of LI.\citep{Cohen}  In contrast, Chodos has devised a new symmetry he calls Light Cone Reflection (LCR), which enlarges the SIM(2) symmetry of VSR.\citep{Chodos2} He shows that one can construct a Lagrangian describing neutrinos that satisfies LCR, in which  $\pm m^2$ neutrino pairs arise naturally. Thus, the Chodos model explicitly requires neutrinos come in tachyon-tardyon pairs having the same magnitude mass, just as we have suggested for two of the three neutrino flavors. While the presence of a third unpaired neutrino would seem to present a problem for the Chodos model, it is natural to imagine that it could be resolved by having a fourth sterile neutrino in addition to the three active ones.

As noted earlier BM in one version of their fits consider this 3 + 1 scenario.  Unfortunately for the Chodos model, their best fit in this case yields $\Sigma m_\nu = 0.06 eV$ and $m_{st}= 0.450 \pm 0.124 eV,$ for which the three active neutrino masses would need to have masses far smaller than the sterile one, a result that seemingly rules out the possibility of two tachyon-tardyon $\pm m^2$ pairs.  Nevertheless, there is a way to achieve consistency with the Chodos model.  The assumption made by BM in doing their fits was that the minimum possible value for $\Sigma m_\nu$ is 0.06 eV based on neutrino oscillation $\Delta m^2$ values.  However, that assumption would be incorrect if some of the neutrino flavor masses are tachyons.  In fact if two rather than one of them is a tachyon there would be two negative mass terms in Eq. 3, and the result for $\Sigma m_\nu$ would be negative under the Chodos model.   Since BM report a ``best fit" value for $\Sigma m_\nu$ right at the lower end ($0.06 eV$) of the assumed allowed region, this fact strongly suggests that the true best fit does indeed lie below 0.06 eV and is possibly negative -- particularly if the likelihood function were continually rising as the value of $\Sigma m_\nu$ descended toward 0.06 eV.  Interestingly, BM do not show a plot of their likelihood function for $\Sigma m_\nu$ in the 3 + 1 case, even though they do show it for the 3 + 0 case.  If BM or other investigators should choose to explore the $\Sigma m_\nu < 0$ region in doing their 3 + 1 fitting,  it would be quite interesting to see if the best fit is in accord with the Chodos model.  Specifically, since the model requires the four neutrinos constitute two $\pm m^2$ pairs, then we should find: $\Sigma m_\nu = -m_{st}.$  We should note, however, that such a result would be equally compatible with a 3 + 3 scenario, where the two additional sterile neutrinos constitute a third $\pm m^2$ pair that would add equal magnitude $\pm$ masses to the right hand side of $\Sigma m_\nu = -m_{st}.$ 

Let us assume the existence of a sterile neutrino mass state $\nu_4$ accompanying the usual 3 active mass states with masses $m_1, m_2, m_3,$ having a standard mass hieracrchy.  It is also assumed that $|m^2_1|>>|\Delta m^2_{23}|$ and $m_1^2<0,$ because otherwise it would be impossible to have a tachyonic $\nu_e$ with $- m_{\nu_e}^2=\mu^2_e >>|\Delta m^2_{23}|.$  Under these assumptions the four flavor state masses can be found from:

\begin{equation}
m^2_{F,i}= \Sigma_1^4 \big[m^2_j|U_{ij}|^2\big] \approx m_1^2 (1-|U_{i4}|^2) +m_4^2|U_{i4}|^2 
\end{equation}

where $i = e,\mu,\tau, st.$  There are many ways Eq. 6 can yield two pairs of flavor states having $\pm m^2$ masses with  $(e,st)$ constituting one pair, and hence satisfying: $m^2_{\nu_e}=-m^2_{\nu,st}.$  In this case we can use the value for $m_{\nu,st}$ that Battye and Moss found in their 3 + 1 best fit to obtain $m_{\nu_e}^2=-0.450^2 eV^2,$ or $\mu_{\nu_e}= 0.450\pm 0.124 eV.$  Unfortunately, however, all solutions with only one sterile neutrino yield too large a degree of active-sterile mixing.  Based on neutrino oscillation data, for example, it is found that $1-\Sigma_1^3 |U_{i,j}|^2<.01,$\citep{Antusch} which is far less than the minimum value possible using Eq. 6 that can be easily shown to be 0.5. However, with $\emph{three}$ sterile neutrinos many solutions exist that have an arbitrarily small degree of active sterile mixing, and which yield three $\pm m^2$ pairs -- see Appendix I.  Thus, the third observation that $\mu_{\nu_e}= 0.450\pm 0.124 eV$ uses the BM 3 + 1 fit result in the context of the Chodos model.  Of the BM fits to the CMB and lensing data in this section and the preceding one, the one in this section is preferred, since the 3 + 0 fit (sect. IIB) cannot be compatible with very small $\Delta m_{ij}^2$ from oscillation experiments.

\subsection{Fine structure in the knee region of the cosmic ray spectrum}
Recall that the original basis of the predicted $E\approx 4.5 PeV$ peak was that cosmic ray protons were hypothesized to decay when $E>E_{knee}=1.695 PeV/\mu_{\nu_e}$ if $\nu_e$ is a tachyon, which is how the 1999 estimate for $\mu_{\nu_e}=0.50 \pm 0.13 eV$ was obtained.  The position of the knee for cosmic ray protons is now known to depend on cosmic ray composition, and for protons it is claimed to be $E_{knee}=4.0\pm 1.0 PeV,$\citep{Apel1} which when substuted in Eq. 1 yields $\mu_{\nu_e} = 0.43 \pm 0.11 eV.$  The observation in support of this result consists of data reported by the Tunka Collaboration,\citep{rio_conf} which is interpreted here as providing further evidence for the predicted $E\approx 4.5 PeV$ peak.  Most earlier cosmic ray experiments exploring the knee region show merely a change in power law, i.e., a knee.  Tunka, however, reports seeing ``remarkable fine structure" in the knee region and at higher energies for their all-particle spectrum -- see Fig. 2 here.  Tunka authors attribute the observed fine structure to a combined source model where cosmic rays around the knee are produced by the group of SN Ia remnants and the extragalactic light component (in accordance with ``dip’' model) arises in the energy region of $10 - 100 PeV.$   

\begin{figure}[h]
\includegraphics[width=100mm]{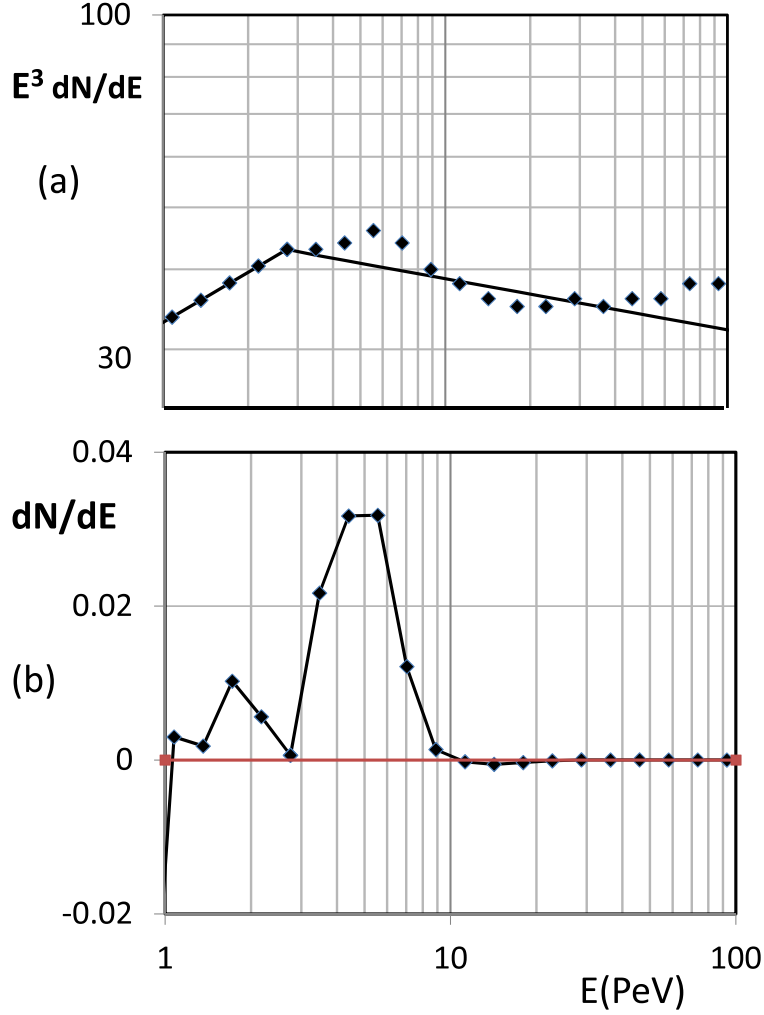}
\caption{(a) Shows the data for the all-particle spectrum flux times $E^3$ extracted from the Tunka paper (reference 13) together with two straight lines representing with a change in the power law from $E^{-2.74}$ to $E^{-3.08}$ at $E \approx 3 PeV$.  These three numerical values were chosen maximize the prominence of the peak near 5 PeV.  The error bars in (a) are smaller than the points plotted.  In (b) we see the excess flux above those two straight lines as a function of energy in PeV.  Lacking error bars, it was not possible to due a true fit, and make a null test of the peak.  Note that only in (a) is the flux multiplied by $E^3$ which accounts for $dN/dE$ vanishing for $E>20 PeV$ in (b).} 
\end{figure}

As noted in Fig. 2 (b), however, the fine structure seen could also be characterized as consisting of a noticeable peak in the range $E\approx 5 PeV$ superimposed on a straight change in power law at about 3 PeV, i.e., the knee.  Assuming the peak is real, it is possible that very good energy resolution is needed in order to see it -- in Tunka's case $\Delta E/E \approx 15\%$ for $E>1PeV$.\citep{Tunka}   Indeed, if we artificially blur the Tunka data in energy corresponding to a resolution of $50\%$ the evidence for any peak largely disappears.  Moreover, as the Tunka authors note, a similar fine structure has been seen above the knee in some other recent experiments, including KASCADE Grande\citep{Apel} and Ice Top.\citep{IceTop} See for example Fig. 8 in ref.~\citep{IceTop} which shows the same dip Tunka sees at about 20 PeV in those two experiments -- a dip that gives rise to the peak seen in our Fig. 2 (b).

\subsection{The second knee in the cosmic ray spectrum}
Another feature in the cosmic ray spectrum known as the `second knee' (a second abrupt increase in spectral softness) has been observed recently in the vicinity of $300 PeV$ by the TAIGA collaboration,\citep{Tunka1} and it was also noted previously by three other experiments at a slightly higher energy (330 PeV).\citep{Bergman}.  The second most abundant nucleus in the cosmic rays is helium, so if we have interpreted the first knee as the threshold for proton beta decay it is reasonable to expect the less prominent second knee as being due to $He^4$ decay via the reaction having a final state $\nu_e$ or $\bar{\nu_e}$ that has the lowest threshold (least negative Q-value):

\begin{equation}
He^4 \rightarrow H^3 + n +e^+ + \nu_e
\end{equation}

Note that the sudden onset of this process results in more and more $A=4$ particles being replaced by lighter ones.  The attribution of the second knee as being due to the onset of $He^4$ decay therefore receives further strong support by the sudden lowering in the mean atomic mass number of cosmic rays that occurs beginning with the energy of the second knee also seen in the TAIGA data.\citep{Tunka1} Applying Eq. 1 with the proton mass replaced by that of an alpha particle, one can express the predicted threshold energy for $He^4$ beta decay in terms of the tachyonic neutrino mass $\mu_{\nu_e}$ and equate it to the observed energy of the second knee:

\begin{equation}
E_{2^{nd}knee}=\frac{m_{He^4}|Q|}{\mu_{\nu_e}}=\frac{80.6 PeV}{\mu_{\nu_e} (eV)}=300\pm 100 PeV
\end{equation}

where the $100 PeV$ uncertainty in the knee position is inferred from the graph in ref.\citep{Tunka1}. Solving Eq. 8 we find that $\mu_{\nu_e} = 0.27 \pm 0.09 eV.$  

\subsection{Neutrino-less double beta decay $0\nu ({\beta\beta})$}
A sixth and final observation comes from an experiment looking for $0\nu ({\beta\beta})$, a very rare process requiring $\nu_e$ to be a Majorana particle.  Recently, solutions of the tachyonic Dirac equation, originally proposed by Chodos, Hauser and Kostelecky\citep{Chodos}, have been studied in the helicity basis, by Jentschura and Wundt leading to a consistent description of a tachyonic spin 1/2 Dirac field.\citep{Jentschura, Jentschura1}  Chang has also worked with a tachyonic form of the Dirac equation, and shown that Majorana solutions can be constructed.\citep{Chang} Thus, both Dirac and Majorana solutions are possible for tachyonic neutrinos.  According to the standard theoretical mechanism involving the exchange of a light exchanged Majorana neutrino, the effective mass of the $\nu_e$ in $0\nu ({\beta\beta})$ can be inferred from the observed half-life of a decaying nucleus.  Moreover, if CP is conserved and $|m_{\nu_e}|> 0.1 eV,$ the effective mass of $\nu_e$ would be the same in double and single beta decay.\citep{Vogel} The relationship between $\nu_e$ mass and half-life in $0\nu ({\beta\beta})$ is:

\begin{equation}
|m_{\nu_e}^2|=\frac{m_e^2}{T_{1/2}G|M^2|}
\end{equation}

where G is the phase space available, $m_e$ is the electron mass and M is the nuclear matrix element.\citep{Rodejohann}  Note however that by Eq. 9 the measured half-life is insensitive to the sign of $m_{\nu_e}^2.$  The most sensitive experiment done as of 2001 reported merely an upper limit for $|m_{\nu_e}|<0.3 - 1.0 eV.$\citep{Heidelburg} However, 
in later papers, with improved statistics and a different method of analysis, the Heidelberg-Moscow (H-M) collaboration published a series of papers reporting an actual value rather than simply an upper limit.\citep{Heidelburg1, Heidelburg2, Heidelburg3}  In 2006, after 13 years of data-taking for the decay of $^{76}Ge,$ they reported a $6.4\sigma$ signal, with a half-life corresponding to $|m_{\nu_e}| =0.32\pm 0.03 eV.$\citep{Heidelburg1} This result has been challenged because of questions whether this rare nuclear process  was in fact observed above background.  In fact, the GERDA experiment,\citep{GERDA} and two others looking for $0\nu ({\beta\beta})$ using $^{138}Xe$\citep{Kamland,EXO} have recently also reported negative results at a level of sensitivity that is claimed to contradict the H-M 2006 positive result.  However, the H-M authors have disputed the fact of a contradiction based on considerations involving energy resolution and insufficient statistics in the negative results.  The GERDA and H-M groups have written ``dueling" letters to the CERN Courier explaining why their result is the correct one.\citep{Courier}  Of course, a resolution of the $0\nu \beta\beta$ controversy in favor of the negative results would not be fatal to our hypothesis, since tachyonic neutrinos need not be Majorana fermions.

\section{Conclusionsand future tests}

We have discussed six observations which are each consistent with the hypothesis that $\nu_e$ is a tachyon having a mass $\mu_{\nu_e} =0.33 \pm 0.024 eV.$    Two of the observations (B and C) rely on fits to the same CMB and lensing data, and they represent different interpretations of that data depending on whether sterile neutrinos exist (C) or not(B). It may be tempting to dismiss the six observations as a random list of unrelated items, but they all would follow directly from the existence of a tachyonic $\nu_e$ and they all suggest the same mass -- one that is consistent with what was originally proposed in 1999,\citep{Ehrlich1} and given further support in a second publication that year.\citep{Ehrlich2}  It may also be true that each observation taken individually has a more mundane explanation, but it should be noted that there are no known unobserved phenomena that should have been observed assuming the claim to be correct, as for example was the case for the original OPERA result, as noted by Cohen and Glashow.\citep{Cohen1} 

Is it possible that the observations yield tachyonic masses due to common systematic errors in the data from which they were derived, as probably was the case in tritium decay experiments, 13 out of 14 of which have yielded $m_{\nu_e}^2<0$ best fit values?\citep{PDG}  The systematic errors involved in most of the six observations, however, are of a completely different character from one another, i.e., there is no reason to believe that possible systematic errors involved with finding non-uniformities in the cosmic background radiation have anything whatsoever to do with systematic errors in looking for $0\nu$ double beta decay, or those in a cosmic ray experiment.  Thus, the possibility of common systematic errors accounting for the six observations yielding a tachyonic mass for $\nu_e,$ let alone the $\emph{same}$ mass does not appear to be realistic.

Corroboration or refutation of the claim of this paper could come in a variety of ways, but a direct measurement of a superluminal speed is of course not one of them.  A time-of-flight experiment to measure the excess above light speed for a $\nu_e$ with $m_{\nu_e}^2 = -0.11 eV^2$ is out of the question, given the energy thresholds of neutrino detectors.  For example, assuming $E=1 MeV,$ one finds $(v/c -1) = 10^{-14},$ so that over a 600 km distance, $\nu_e$ would outrace a photon by an immeasurable $2\times 10^{-17} s.$   There are, however, more promising tests that could test the claim that $\mu_{\nu_e} =0.33 eV$ including three listed below. 

\begin{enumerate}

\item $\textbf{Further evidence of a 4.5 PeV peak in cosmic ray data}$\\  
Even without any well-established sources of PeV cosmic rays, one could look for excess numbers of cosmic rays in specific small regions of the sky for E near $4.5 PeV,$ as suggested by the results in ref.~\citep{Ehrlich4}.  Cosmic ray researchers should also examine data in the PeV range within several degrees of Cygnus X-3.   In the 1999 paper\citep{Ehrlich2} making this claim a peak was seen in an energy histogram for events in a narrow ($\Delta \phi=2.5\%$) phase window associated with the source's $T= 4.79 h$ period, using as background the counts in the remaining $97.5\%.$   This made for an ideal choice of background in a null test if Cygnus is not a source.  In fact, the $\emph{only}$ excess above background found was a $6\sigma$ peak centered on $E=5 PeV.$  This claim has been viewed skeptically because of an earlier negative search result for signals from Cygnus X-3 in a very high statistics experiment.~\citep{CASA}  However, that experiment could not possibly have seen a $4.5 PeV$ signal associated with Cygnus X-3, given that only $0.09\%$ of its events had $E>1.175 PeV.$  

In order to test the claim one must use a proper phase window selection, which requires that there be a sizable fraction (and number) of events in the $4.5 PeV$ region, so that a statistically significant sharp peak could be seen in a phase distribution for events within a few degrees of Cygnus X-3.  Equally important, the phase of each event needs to be calculated using a highly accurate ephemeris.  For example, suppose one had 10 y worth of data, i.e., $N=18,000$ Cygnus X-3 periods.  In that case, the (time varying!) period for Cygnus X-3 would need to have an uncertainty less than $\Delta \phi/(2N)=0.00007\%$ in order to see a sharp phase peak in a phase histogram.  The 1999 claim of a 4.5 PeV neutron peak from Cygnus X-3 appears never to have been tested based on a citation search.  Moreover, of the three tests proposed here, only this one could be made in a relatively short time using archived data.  It would be especially useful to look for a $4.5 PeV$ peak in cosmic rays from Cygnus X-3 in an experiment where the hadronic nature of neutral particles possibly from that source could be established,\citep{Marshak1, Marshak2} and/or during times when major flares occur, when radio emissions have been observed to increase a thousandfold.

\item $\textbf{A direct measurement of $m_{\nu_e}^2$ in a tritium beta decay experiment}$\\ 
According to the KATRIN collaboration, they should be able to discover the $\nu_e$ mass if it exceeds 0.30 (0.35) eV at a 3 (5) $\sigma$ level).\citep{KATRIN}  The ability to discover the $\nu_e$ mass if it is a tachyon with $\mu_{\nu_e} = 0.33 eV$ need not be the same as for a tardyon having the same magnitude mass.  As can be seen from the Kurie plots in Fig. 3 near the endpoint region the difference from the linear ($m_{\nu_e}=0$) curve is far greater when $m_{\nu_e}^2>0$ than $m_{\nu_e}^2<0,$ suggesting a greater likelihood for detecting $m_{\nu_e}^2<0,$  when in fact $m_{\nu_e}=0,$ in agreement with the possible bias yielding $m_{\nu_e}^2<0$ values in 13 out of 14 tritium experiments.

KATRIN has run over 12,000 simulated spectra assuming $m_{\nu_e}=0,$ so as to test their fitting procedures.\citep{Riis}  In these simulations they find that the distribution of the best fit masses  is well described by a symmetric Gaussian function centered on zero out to around $\pm 2\sigma,$ which would suggest a similar sensitivity for both signs $\pm m_{\nu_e}^2.$  The symmetry of the distribution, however, may not hold beyond $\pm 2\sigma.$  In fact, the distribution from the 12000+ simulations does favor $m_{\nu_e}^2>0$ over $m_{\nu_e}^2<0$ beyond around $\pm 2\sigma.$   Thus, in a histogram of their simulated results for $m_{\nu_e}^2>40 eV^2,$  one finds 38 cases compared to only 19 cases for $m_{\nu_e}^2<-40 eV^2$ -- see Fig.1 in ref.~\cite{Riis}.    Based on the preceding discussion, while KATRIN may have the sensitivity to discover a $\nu_e$ mass above 0.35 eV at a $5\sigma$ level, it may be unlikely to be able corroborate the claim of this paper that $\mu_{\nu_e} =0.33 eV$ at that same level.  Nevertheless, even a $3-4\sigma$ KATRIN result consistent with $\nu_e$ being a tachyon with the specific value $\mu_{\nu_e} = 0.33 eV$ would be quite interesting.  

\begin{figure}[h]
\includegraphics[width=80mm]{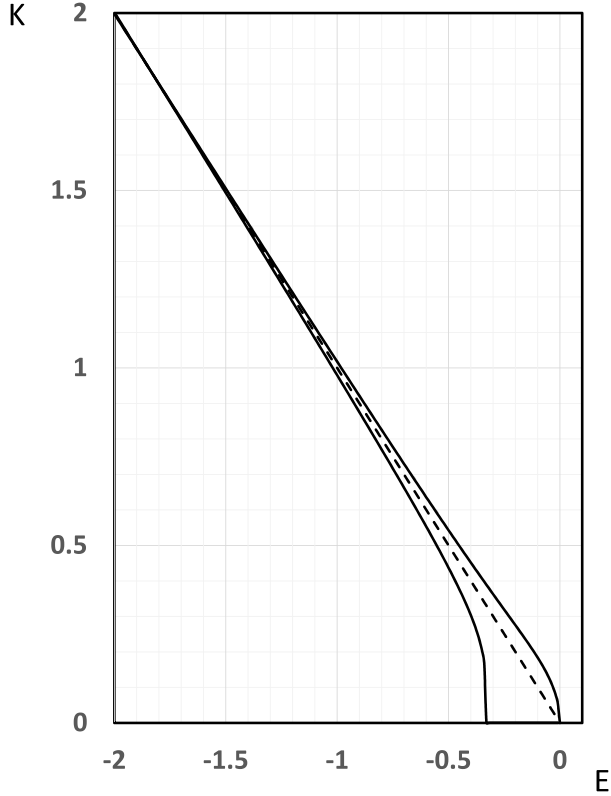}
\caption{The curves are plots of the Kurie function $K=C\large[\sqrt{E_0+E)^2-m_\nu^2}\times (E_0+E)\large]^{1/2}$ in the region within 2 eV of the endpoint $E_0$ of the beta spectrum.  The linear dashed curve corresponds to a massless $\nu_e,$ and the upper/lower solid curves correspond to a tachyonic and tardyonic $\nu_e$ having a $\mu_{\nu_e}$ or $m_{\nu_e} = 0.33 eV$ respectively.  The constant C for each curve has arbitrarily been chosen so that they have the same height at $E=-2 eV.$} 
\end{figure}

\item $\textbf{The neutrino burst from a core-collapse galactic supernova}$\\  
If the core-collapse were to exhibit time variations in the neutrino emissions on the scale of a few milliseconds as suggested by recent two-dimensional simulations, and if those time structures were seen for the arriving neutrinos in a wavelet analysis,\citep{Ellis} such an observation would disprove the hypothesis.   Conversely, if those ms-scale time variations were smeared out by $E_\nu$-dependent travel times, it should be possible to determine the $m_\nu^2$ that best unsmears the data, i.e., reproduces the time distribution at the source, by subtracting from each $\nu_e$ arrival time the quantity $\Delta t= t_0 m_\nu^2/2E_\nu^2,$ where $t_0$ is the light travel time from the supernova.   The analysis in ref.~\citep{Ellis} suggests that this might result in an $m_\nu^2$ uncertainty $\Delta m_\nu^2 \approx \pm 0.016 eV^2$ which should easily permit a verification of a $\mu_{\nu_e} =0.33 eV$ or $m_{\nu_e}^2 =-0.11eV^2$ tachyon.\\  

An alternate analysis to test the hypothesis would be based on a measurement of the slope of the leading edge of the neutrino pulse.  Let us define the leading edge of the $\emph{emitted}$ neutrino pulse to comprise the first 0.01 seconds.  Consider two leading edge neutrinos emitted simultaneously having energies $E_1$ and $E_2>E_1,$ with $E_1$ being the lowest energy that can reliably distinguished from background. Based on relativistic kinematics, the difference in their arrival times in the detector would be: 

\begin{equation}
\Delta t = t_2 - t_1= \frac{t_0 m_\nu^2}{2}\big(E_1^{-2} - E_2^{-2}\big)
\end{equation}

For another supernova at the same distance as SN 1987a (168kly), if we let $E_1=5 MeV$ and $E_2 = 50 MeV,$ we would find that the lower energy neutrino arrived $0.01s$ $\emph{before}$ the higher energy one.  Thus, on a plot of $1/E^2$ vs arrival time $t$ we should be able to deduce $m_{\nu_e}^2$ from the slope of the leading edge of the pulse of arriving neutrinos, assuming the correctness of supernova models that show the pulse rise time at the source for $\nu_e$ to be much less than $0.01s.$\citep{Hudepohl}  This assertion also assumes a standard neutrino mass state hierarchy, and a detector with better than millisecond timing that observed thousands of $\nu_e$ from a supernova at 168 kly.  In the unlikely event the neutrino mass eigenstates have a highly nonstandard hierarchy, the data might even reveal the arrival of each mass state separately.  This certainly would be the case if the masses are as large as suggested in several SN 1987A analyses.\citep{Ehrlich5, Ehrlich6} 

\end{enumerate}
\begin{acknowledgments}
The author thanks Alan Chodos, Ulrich Jentschura and Ngee-Pong Chang for helpful comments.
\end{acknowledgments}
\newpage
\section*{Appendix I:  Three sterile neutrinos and the Chodos model}

The 6 x 6 neutrino mixing matrix can be expressed in this form:\citep{Xing}

\begin{equation}
\left(
\begin{array}{c}
\nu_e\\
\nu_\mu\\
\nu_\tau\\
\nu_x\\
\nu_y\\
\nu_z\\
\end{array}
\right)=\left(
\begin{array}{cccccc}
&&|&  & & \\
  U_{ij} &&  |&  & S_{ij} & \\
--& -- &  |& -- & -- & --\\
 &  & |&  & & \\
  -S_{ji} && | & &V_{ij} &\\
 &  &  |& & &\\
\end{array}
\right) =
\left(
\begin{array}{c}
\nu_1\\
\nu_2\\
\nu_3\\
\nu_4\\
\nu_5\\
\nu_6\\
\end{array}
\right)
\end{equation}

where the entries for $U_{ij}$ are based on the standard 3 x 3 matrix of mixing parameters found from the 3 measured mixing angles for the active neutrinos, the $S_{ij}$ designates the 9 mixing parameters between the 3 active and 3 sterile states, and the $V_{ij}$ are the mixing parameters between the mass states $\nu_4, \nu_5, \nu_6.$  In general, the expressions for the $S_{ij}$ in terms of the 9 active-sterile mixing angles is messy,\citep{Xing} but in the special case where we want to have all $|S_{ij}|<<1$ so as to keep active-sterile mixing very small and have ``minimal non-unitary,"\citep{Antusch1} we have the simple result: $S_{ij}=sin \theta_{ij}, i=1,2,3, j=4,5,6$.  

We can use Eq. 8 to find expressions for the flavor state masses in the usual way, i.e., extending the right hand side of Eq. 4 to sum over six mass state masses $m^2_1\cdots m^2_6.$  If we again assume $|m^2_1|>>|\Delta m^2_{23}|$  there would be a total of 16 adjustible parameters: $m^2_1, m^2_4, m^2_5, m^2_6,$ the nine $S_{ij},$ and the three mixing angles between the sterile mass states.  Given such a large number of adjustible parameters, many solutions exist for the 6 flavor state masses (3 active + 3 sterile) that result in these three $\pm m^2$ Chodos pairings:\citep{solutions} 

\begin{equation}
m^2_{\nu_e}=-m^2_{\nu,Z}  \hspace{.25in} m^2_{\nu,\mu}=-m^2_{\nu,\tau} \hspace{.25in} m^2_{\nu,X}=-m^2_{\nu,Y}
\end{equation}

Thus, assuming a mass $m_Z= 0.450 eV,$ as suggested by the BM 3+1 fit, solutions for $m^2_{\nu_e}$ can be found satisfying $\mu_{\nu_e}= 0.450\pm 0.124 eV,$ which are consistent with the Chodos model, and satisfy all empirical constraints, i.e., the measured three mixing angles, and very little active-sterile mixing.

\huge{Wenn Sie ${m^2_\nu}_e$ zu messen seien Sie ehrlich!}\\
$\rho_{rad}=\frac{Energy}{Volume}=\rho_\gamma (T) + \rho_\nu(T)N_\nu$\\
$t \rightarrow -t$ and $E \rightarrow -E$\\
$\nu_e,\nu_\mu,\nu_\tau$\\
$\rho_\nu=n_em_e+n_\mu m_\mu+n_\tau m_\tau=n_\nu\Sigma m_\nu$\\    $(1)\nu_e=\bar{\nu}_e?\hspace{0.2in}$
$(2){m^2_\nu}_e<0?$\\
$(3)<m_{\beta\beta}>=<m_\beta>?$\\
$|\Delta N_\nu|<0.3$\\
$0.36 \pm 0.041 eV$\\
$\Sigma m < 0$\\
$n\rightarrow p\rightarrow n\rightarrow p\rightarrow n\rightarrow p\rightarrow \cdots $\\
$\nu \rightarrow \nu e^+e^-$\\
$\Delta m_{13}^2=0.00232 eV^2$\\
$T_\nu=(\frac{4}{11})^{1/3}\approx 1.9 K$\\
$\mu=\sqrt{-m^2}$\\
$\mu=0.50\pm 0.13 eV$\\
$m^2=E^2-p^2$\\
$\frac {dE}{dv}$\\
$p \rightarrow n+e^++\nu$\\
$\bar{\nu}+p \rightarrow n+e^+$\\
$E_\nu>0$\\
$E_\nu<0$\\
\begin{equation}
E'=\gamma (E - \beta p)\\
\end{equation}
\begin{equation}
\beta > E/p < 1 \rightarrow E' <0\\
\end{equation}
\begin{equation}
\beta=\sqrt{1-1/\gamma^2}\approx\frac{m^2}{2E^2}\\
\end{equation}
\begin{equation}
\Delta m_{12}^2, \Delta m_{23}^2, \Delta m_{13}^2, \Delta m_{14}^2? \cdots
\end{equation}
\begin{equation}
\Delta t= t_0 m_\nu^2/2E_\nu^2 
\end{equation}
\begin{equation}
\Sigma m
\end{equation}
\begin{equation}
\nu_1, \nu_2, \nu_3, \nu_4? \cdots
\end{equation}
\begin{equation}
\nu_e, \nu_\mu, \nu_\tau, \nu_{st}?
\end{equation}
\begin{equation}
\rho_{rad}=\rho_\gamma+\rho_\nu=\frac{\pi^2}{15}T_\gamma^4 \rho_\gamma+\rho_\nu=\frac{\pi^2}{15}T_\gamma^4 +N_{\nu}\frac{7\pi^2}{120}T_\nu^4
\end{equation}
\begin{equation}
N_\nu=3.05
\end{equation}
\begin{equation}
\Delta N_\nu=N_\nu - 3.046
\end{equation}
\begin{equation}
\rho_\nu=0.2271N_\nu\rho_\gamma
\end{equation}
\begin{equation}
\rho_\gamma=\frac{\pi^2}{15}T_\gamma^4 
\end{equation}
\begin{equation}
\rho_{rad}=(1+0.2271N_\nu)\rho_\gamma
\end{equation}

\begin{thebibliography}{}
\bibitem{solutions} For example, given an arbitrary  choice of the 3 mixing angles defining the $V_{ij}$ and the 3 defining the $S_{ij},$ we may use Eqs. 8 to define 3 linear equations in the unknowns $m^2_1, m^2_4, m^2_5, m^2_6.$  A fourth linear equation is defined by $m_Z= 0.450 eV.$  If the mixing angles were all chosen at random the four linear equations would be neither linearly dependent nor inconsistent, and would have a unique solution.
\end{thebibliography}
\end{document}